\documentclass[aps,prb,reprint,groupedaddress,showpacs,amsmath,amssymb,superscriptaddress]{revtex4-1}
\usepackage{graphicx}
\usepackage{caption}
\usepackage{dcolumn}% Align table columns on decimal point
\usepackage{bm}% bold math
\usepackage{float}
\usepackage{url}
\usepackage{amssymb}
\usepackage{subfigure}
\usepackage{calc}
\usepackage{latexsym}
\usepackage{epsf}
\usepackage{epsfig}
\usepackage{notoccite}
\usepackage{euscript}
\usepackage{hyperref}
\usepackage{xcolor}
\usepackage{sidecap}
\usepackage[english]{babel}
\usepackage[autostyle]{csquotes} %Allows for appropriate usage of quotation marks without difficulty

\begin{document}

\title{Spin Canting and Orbital Order in Spinel Vanadate Thin Films}

 \author{Christie Thompson}
 \affiliation{Materials Science and Engineering Program, FSU, Tallahassee, FL 32310, USA}
 \affiliation{National High Magnetic Field Laboratory, FSU, Tallahassee, FL 32310, USA}
\author{Dalmau Reig-i-Plessis}
 \affiliation{Department of Physics, University of Illinois at Urbana-Champaign, USA}
 \author{Lazar Kish}
  \affiliation{Department of Physics, University of Illinois at Urbana-Champaign, USA}
 \author{Adam A. Aczel}
 \affiliation{Neutron Scattering Division, Oak Ridge National Laboratory, Oak Ridge, TN 37831, USA}
 \author{Biwen Zhang}
  \affiliation{National High Magnetic Field Laboratory, FSU, Tallahassee, FL 32310, USA}
 \affiliation{Department of Physics, FSU, Tallahassee, FL 32306, USA}
 \author{Evguenia Karapetrova}
 \affiliation{Advanced Photon Source, Argonne National Laboratory, Argonne, Illinois 60439, USA}
\author{Gregory J. MacDougall}
 \affiliation{Department of Physics, University of Illinois at Urbana-Champaign, USA}
\author{Christianne Beekman}
 \affiliation{National High Magnetic Field Laboratory, FSU, Tallahassee, FL 32310, USA}
  \affiliation{Department of Physics, FSU, Tallahassee, FL 32306, USA}
\date{\today}

\begin{abstract}
\noindent We report on the epitaxial film growth and characterization of CoV$_2$O$_4$, a near-itinerant spinel vanadate, grown on (001) SrTiO$_3$. The symmetry lowering of the unit cell from cubic in the bulk to orthorhombic in the films results in dramatic differences in the magnetic anisotropy compared to bulk, as determined from structural and magnetic characterization. Bulk cubic CoV$_2$O$_4$  has been found to defy predictions by showing orbital degeneracy seemingly lasting to very low temperatures, with only small anomalies in magnetization and neutron experiments signaling a possible spin/orbital glass transition at T = 90 K. In epitaxial thin films presented in this paper, structurally tuning the CoV$_2$O$_4$ away from cubic symmetry leads to a completely different low temperature non-collinear ground state. Via magnetization and neutron scattering measurements we show that the 90 K transition is associated with a major spin reorientation away from the ferrimagnetic easy axis [001] to the [110] direction. Furthermore, the V-spins cant away from this direction with extracted perpendicular moments  providing evidence of a larger canting angle compared to bulk. This result indicates that compressive strain pushes the system deeper into the insulating state, i.e., away from the localized - itinerant crossover regime. 
\end{abstract}
%\pacs{78.20.-e,78.20.Ci,78.40.Fy}
\maketitle

%\section{INTRODUCTION}
\noindent Magnetic oxides composed of a frustrated magnetic network of 3d transition metals, in which orbital, spin and structural degrees of freedom are strongly coupled \cite{Lee_rev}, have been an active playground for researchers due to the promise of finding routes to new behaviors. In recent years the focus has been more and more on geometrically frustrated systems in which spin-spin interaction result in macroscopically degenerate ground state manifolds \cite{Venderbos,moessner,ramirez1}, lots of entropy, a large density of states that can be manipulated, and emergence of unusual low temperature properties when perturbations are applied. Furthermore, localized spin and itinerant electron behavior are strongly coupled in geometrically frustrated systems \cite{lacroix} leading to spin liquid behavior \cite{nakatsuji} and other exotic phases \cite{hanasaki,iguchi,kumar} as observed in pyrochlores.  %Near-itinerant spinel vanadates in particular display collective transitions that often involve changes in multiple degrees of freedom \cite{Lee_rev}. It has been shown that the competition between orbital order and itinerancy \cite{ma} triggers many interesting phenomena such as metal-insulator transitions \cite{imada}, colossal magnetoresistance \cite{ramirez}, superconductivity \cite{JACE:JACE3317} and multiferroicity \cite{cheong, zhang, Myoung, lin,Dey,Mufti}. 

Spinel vanadates, in which itinerancy and frustration can be controlled via manipulation of the V-V distance \cite{huang, ma, kaur, kisma,kisma2,kiswandhi}, are poster materials for orbital physics in frustrated antiferromagnets. They have been intensely studied to gain a better understanding on how orbital order can help relieve spin degeneracy. Vanadates (AV$_2$O$_4$) with non-magnetic atoms on the A-site (A = Zn, Cd, Mg \cite{zhang,LeeZnV,Ueda,Mamiya,Nishiguchi,Wheeler,Lee_rev}) show two successive transitions, first a structural transition that leads to orbital order, and at lower temperature, a transition to an antiferromagnetic state. It has been shown before that these vanadates can have different structural distortions and thus different low temperature orbital states \cite{Lee_rev}. Vandates with magnetic atoms on the A-site (A = Fe, Mn) show multiple structural phase transitions as a function of temperature, eventually leading to non-collinear and orbitally ordered ground states \cite{greg2012}. The origin of the spin canting transitions in vanadates is still under debate \cite{Lee}, although recently hyperfine magnetic interactions have been indicated as a possible cause in bulk FeV$_2$O$_4$ \cite{Myoung}. CoV$_2$O$_4$ has garnered extra attention in the study of the vanadates, because it is the closest known material to an identified localized-itinerant cross-over\cite{kisma,kaur,blanco}. Experimentally, bulk CoV$_2$O$_4$ defies predictions by showing orbital degeneracy seemingly lasting to very low temperatures. Only recently a weak spin canting and a first order structural transition associated with an orbital glass transition has been identified at T = 90 K at the edge of detectability \cite{reig, kobinerai}. The proximity to itinerancy has been indicated as the cause for the difficulties in observing these transitions reproducibly.

In contrast to the weak effects seen in cubic bulk samples, here we show that orthorhombic CoV$_2$O$_4$ thin films grown onto SrTiO$_3$ substrates demonstrate unmistakable signatures of spin canting and structural effects that indicate long-ranged orbital order. From these experimental results we conclude that a symmetry lowering due to application of in-plane compressive strain can drive the system deeper into the insulating state. This shows that structural tuning via epitaxial strain is a viable knob for manipulation of non-collinear spin states, orbital states, and itinerancy in spinel vanadates. %Very few groups are investigating strain engineered properties in thin film spinel and pyrochlore vanadates, so this work represents the beginning of a productive effort into a relatively unexplored area of frustration research.

%\section{EXPERIMENTAL DETAILS}
High quality thin films of CoV$_2$O$_4$ have been grown onto (001) SrTiO$_3$ substrates via pulsed laser deposition using a home-made pressed pellet of CoV$_2$O$_6$. The films were grown in a background pressure of P = 1x10$^{-7}$ Torr, while the substrate was held at 600 $^{\circ}C$. The laser fluence was $\sim$ 0.6 J/cm$^2$ and repetition rate of the laser was 1 Hz.  Atomic force microscopy (AFM) measurements were performed on the SrTiO$_3$ substrates before growth, and on the  CoV$_2$O$_4$ films after growth was completed, using an Asylum MFP-3D system in tapping mode. AFM measurements indicate film surfaces are smooth (root-mean-square roughness $\sim$ 2.5 nm) with small grains visible on the surface (see supplemental material \cite{suppmat}). Film thicknesses were determined with x-ray reflectivity measurements or from transmission electron microscope (TEM) images. Representative TEM images of the film-substrate interface are shown in the supplemental material \cite{suppmat}.
Structural characterization was performed with four-circle synchrotron x-ray diffraction (XRD) measurements at the Advance Photon Source at Argonne National Laboratory, on beamline 33-BM-C \cite{KARAPETROVA201152} ($\lambda$ = 0.77504 nm). Magnetization measurements in an applied field of H = 0.1 T as a function of temperature were performed in a Quantum Design magnetic properties measurement system (MPMS). Elastic neutron scattering experiments were performed at the HB-1A beamline at the High Flux Isotope Reactor at Oak Ridge National Laboratory (HFIR-ORNL) (see supplemental material for further details) \cite{suppmat}. %Neutron experiments were performed at the HB1-A beamline at the High Flux Isotope Reactor at Oak Ridge National Laboratory (HFIR-ORNL).

%\section{RESULTS AND DISCUSSION}
Synchrotron XRD measurements were performed to determine the structure, the epitaxy, and the sample to sample variations in crystallinity. The lattice mismatch between CoV$_2$O$_4$ (bulk: 8.4 $\dot{A}$) and two unit cells of SrTiO$_3$ (2x3.905 = 7.81 $\dot{A}$) is $\sim$7.5$\%$, which is relatively large, and should lead to compressive strain assuming that one film unit cell grows on a block of two by two substrate unit cells.  Measurement of the ($hkl$) positions of six or more reflections were used to determine the orientation matrix of each film \cite{Busing}, which yields typical lattice parameters $a$ = 8.36(2) $\dot{A}$, $ b$ = 8.24(5) $\dot{A}$ and $c$ = 8.457(3) $\dot{A}$ ($\alpha$ = $\beta$ = $\gamma$ = 90$^{\circ}$), showing an orthorhombic unit cell for the films. X-ray $\Phi$-scans through $(h0l)$ and $(hhl)$ film reflections showed no indication of peak splitting, confirming the single crystal nature of the samples. It is interesting to note that, despite the large difference in lattice parameters, our films still exhibited cube on cube epitaxy, i.e., the major crystallographic axes of film and substrate are aligned \cite{Budai} (see supplemental material for more details \cite{suppmat}). Furthermore, the lattice parameters are such that we can conclude that the films display asymmetric compressive in-plane strain (along $a$ $\sim$ 0.5$\%$, along $b$ $\sim$ 1.5$\%$).

\begin{figure}[t]
\centering
\includegraphics[width= 2.5in,keepaspectratio]{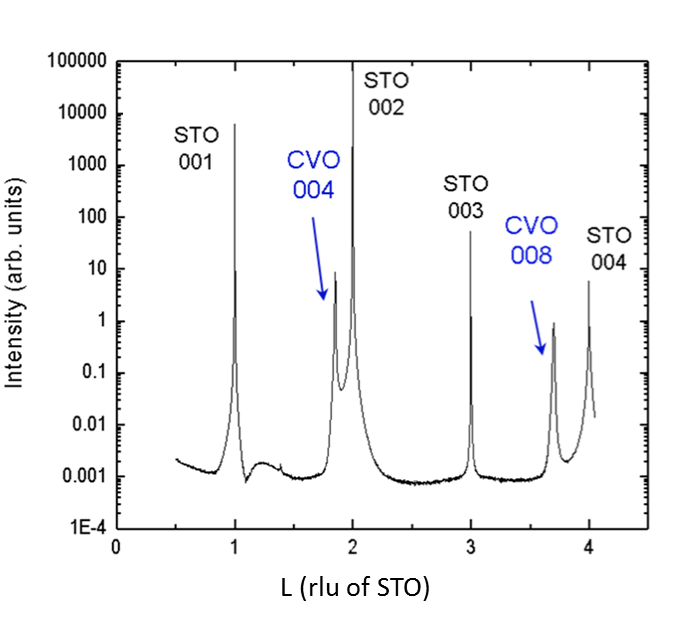}\vskip-0.1in
\caption{\label{char} (Color online) XRD scan of the 00L peaks of a 100 nm thick CoV$_2$O$_4$  film and the SrTiO$_3$ substrate. The film and substrate peaks are indicated with CVO and STO, respectively. }%b) Representative TEM image of a substrate film interface. The scale bar is 5 nm. As indicated by the arrow the $c$-axis points out of the film plane.}
\end{figure}

The 00L scan of a representative sample is presented in Fig. \ref{char}. Both the film and substrate peaks are prominently visible in the L-scan supporting the conclusion on the epitaxial relationship with the substrate.
One film (which we will refer to as ''the cubic polymorph'') showed a structure with the $ c$-axis very close to the bulk cubic lattice parameter ($a$ = 8.34 $\dot{A}$, $b$ = 8.30 $\dot{A}$, $c$ = 8.39 $\dot{A}$) with clear signatures of peak splitting indicating multiple structural domains were present (data not shown). The preparation of this one sample differed from the others, in that the fluence during growth was a little lower, indicating that growth parameters are crucial in structural control of thin films of spinel vanadates. Strikingly, the slight differences in structure and crystallinity between the cubic polymorph and the more orthorhombic films have dramatic consequences for the magnetic properties of the sample as will be discussed next.

\begin{figure}[t]
\centering
\includegraphics[width= 2.5 in,keepaspectratio]{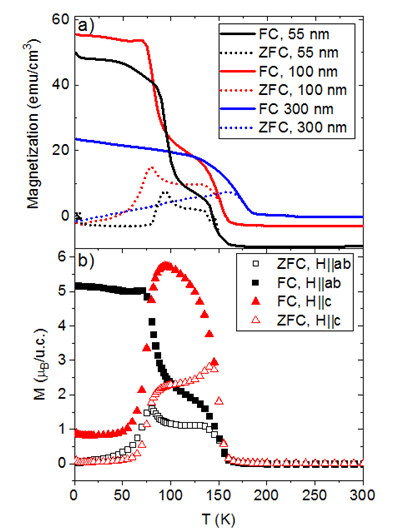}
\caption{\label{MvsT} (Color online) a) M vs. T for a 55 nm orthorhombic film (black curve), a 100 nm orthorhombic film  (red curve), and for the 300 nm cubic polymorph film (blue curve); each are measured while warming in H = 0.1 T after cooling the sample in zero field (ZFC), and after cooling the sample in the measurement field (FC). The field is applied in the film plane, along $a$ or $b$. b) M vs. T for the 100 nm thick film measured in H = 0.1 T (red triangles: H applied out-of-plane of the film, along $c$-axis; black squares: H applied in the film plane along $a$ or $b$),  ZFC open symbols and FC closed symbols. }
\end{figure}

In Fig. \ref{MvsT}a, magnetization as a function of temperature is shown. All measurements were taken in H = 0.1 T while warming. The solid curves represent measurements taken after cooling the sample in $H$ = 0.1 T (FC); the dashed curves represent measurements taken after cooling the sample in zero field (ZFC). Representative curves are shown for two orthorhombic samples (confirmed via XRD measurements) with different film thicknesses (red: 100 nm and black: 55 nm) and for the cubic polymorph (blue: 300 nm). All films with the orthorhombic structure, with thicknesses ranging from 55 - 400 nm, display a ferrimagnetic phase transition at $T \sim$ 150 K similar to bulk samples. However, the orthorhombic films reproducibly show a very obvious second magnetic phase transition at $T$ = 90 K. In bulk this transition has been difficult to observe reproducibly and is associated with a small first order cubic-to-tetragonal structural distortion and spin canting on the pyrochlore V-sublattice (canting angles: 5$^{\circ}$ \cite{reig} and 20$^{\circ}$ \cite{kobinerai} have been reported for bulk polycrystalline CoV$_2$O$_4$, and 11$^{\circ}$ for single crystalline Co$_{1.3}$V$_{1.7}$O$_4$ \cite{kobinerai}). Note, the cubic polymorph  (blue curve in Fig. \ref{MvsT}a) does not show a second phase transition at 90 K. The appearance of the 90 K magnetic transition thus directly ties to the changed unit cell parameters away from cubic symmetry.

Focusing on the 100 nm thick film, in Fig. \ref{MvsT}b magnetization as a function of temperature with H = 0.1 T applied in and out of the film plane (black squares and red triangles, respectively) are compared. The out of plane measurement shows a larger increase in magnetization at 150 K, which is associated with the transition to a collinear ferrimagnetic state. This indicates that the c-axis direction (out of plane)  is the easy axis in the ferrimagnetic state. Surprisingly, this shows that in the orthorhombic structure the single ion anisotropy wins out over the shape anisotropy of the film (which would prefer to have the moment in the plane of the film); this was also observed in the 55 nm film. Furthermore, the large decrease (increase) at the magnetic phase transition at 90 K when the field is applied along the $c$-axis (along the $a$- or $b$-axis) indicates that there is a significant reorientation of the magnetic moment associated with this transition in which most of the moment rotates from the $c$-axis to an in plane direction. This implies that a change in the local symmetry of the sites takes place at 90 K, which affects the strength of the single ion anisotropy, and that the shape anisotropy wins out over the single ion anisotropy.

\begin{figure}[h]
\centering
\includegraphics[width= 2.5in,keepaspectratio]{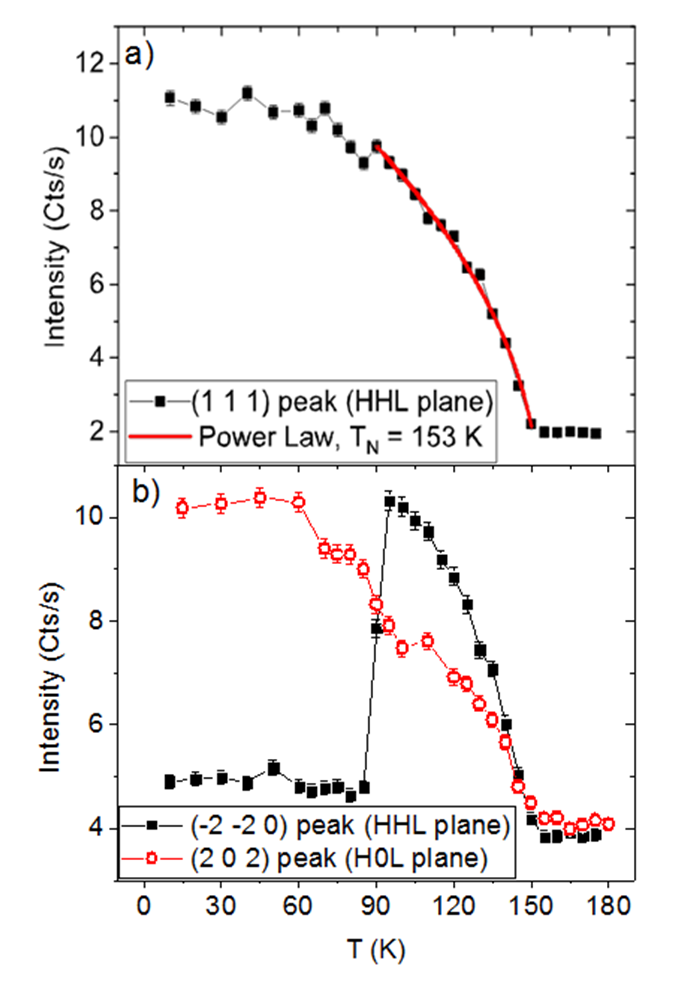}\vskip -0.1in
\caption{\label{neutron111}(Color online) a) (111) Bragg peak intensity as a function of temperature. The red line is a power law fit. b) ($\bar{2}$$\bar{2}$0) (black squares), and (202) (red open circles) Bragg peak intensities as a function of temperature.    }
\end{figure}
\begin{figure}[h]
\centering
\includegraphics[width= 2.5in,keepaspectratio]{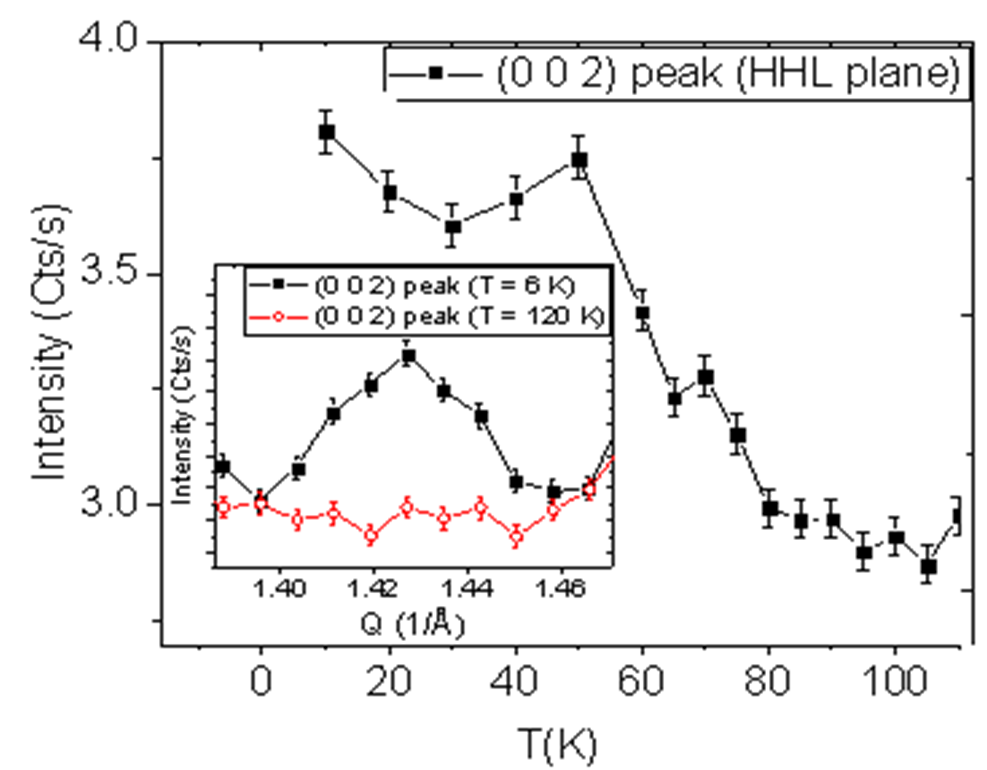}
\caption{\label{neutron002}(Color online)  (002) Bragg peak intensity as a function of temperature. Inset: radial scan of the (002) Bragg peak at 6 K (black squares) and at 120 K (red circles).   }
\end{figure}
Elastic neutron scattering studies in zero applied magnetic field on a $\sim$ 400 nm thick CoV$_2$O$_4$ film grown on a 10 x 10 mm$^2$ SrTiO$_3$ substrate are presented in Figs. \ref{neutron111} and \ref{neutron002}. The (111), ($\bar{2}$$\bar{2}$0), (202), and (002) peaks were monitored as a function of temperature. The onset of the (111) ( Fig. \ref{neutron111}a ) peak can be fit with a power law to reveal the ferrimagnetic phase transition at T$_N$ = 153 K, which is in good agreement with the magnetization data. Strikingly, a clear difference between the ($\bar{2}$$\bar{2}$0) and (202) order parameter curves is observed (see  Fig. \ref{neutron111}b). Note, neutrons probe spin directions perpendicular to the direction of the scattering vector $Q$ and are not sensitive to components of the magnetic moment that are parallel to $Q$. Thus, the increase of both the (202) and ($\bar{2}$$\bar{2}$0) peaks below 150 K, and a faster growth of the ($\bar{2}$$\bar{2}$0) order parameter, confirms that the initial direction of magnetization in the ferrimagnetic state is collinear with the c-axis, i.e., the easy axis. The large drop of the ($\bar{2}$$\bar{2}$0) order parameter at 90 K indicates that the Co-spins are reorienting themselves into the plane of the film when the sample is cooled below 90 K. Fig. \ref{neutron002} shows the (002) order parameter as a function of temperature, which clearly shows that the (002) reflection appears below the 90 K magnetic phase transition (the inset  shows radial scans of the (002) Bragg peak at 120 K and at 6 K.). The (002) is a structurally forbidden peak traditionally associated with spin-canting on the V-sites \cite{reig}. These results strongly imply that in our films the 90 K transition is associated with the V-spin canting away from the Co-spin direction also. 

Based on the experimental evidence presented we can construct a picture of the magnetic properties of CoV$_2$O$_4$ films under compressive strain. %Examining the intensity changes we can determine the canted portion of the V moment using either the low temperature portion of the (002) or the  (-2-20) reflections, and we can determine the Co moment along the [110] using the intensity of the (202) reflection. 
The drop in intensity of the ($\bar{2}$$\bar{2}$0) peak, and the subtle increase in the (202) peak intensity, ambiguously leads to the conclusion that Co$^{2+}$ spin points in the [110] direction below 90 K. Furthermore, the low temperature data presented in  Figs. \ref{neutron111} and \ref{neutron002} can only be explained if we assume a single domain picture for the magnetization, i.e., this implies that the in-plane symmetry is broken in strained thin films. The direction of the vanadium spins is less clear. Previous work on bulk samples suggests that vanadium spins are mostly aligned antiparallel to A-site spins in ferrimagnetic spinels, canting away below an orbital transition \cite{reig,kobinerai}. The finite intensity we observe at the (002) position strongly implies we are seeing a similar canting in a way that breaks the glide plane symmetry of the Fd$\bar{3}$m spacegroup. However, we found the (200) reflection to be absent, which provides clues as to how the V- spins cant. In particular, we have confirmed that any model wherein V$^{3+}$ spins cant towards local $<$111$>$ directions, similar to what is seen in other vanadates \cite{greg2012}, is incapable of explaining our results. Based on the idea that the shape anisotropy wins out at low temperatures, we adopt a planar model, wherein the Co spins point along the [110] direction and the V spins cant away from the [$\bar{1}$$\bar{1}$0] direction but remain in the film plane. This model is consistent with the observed scattering and allows us to extract moment sizes at each atom. The intensity of the (002) and (220) reflections depends only on the canted portion of the V spins, the intensity of the (202) reflection arises from the Co moments, and the intensity of the (111) reflection contains contributions from both the Co and V moments.% with the (0 0 2) and (2 2 0) reflections dependent only on the canted portion of the V spins, and the (2 0 2) reflection dependent only on the Co moments; the (1 1 1) reflection contains contributions from both the Co and V moments.

Based on the normalized intensity of the (202) reflection we find a Co moment of 3.14 ($\pm$ 0.48) $\mu_B$/Co. As expected from our magnetization data and from observations by others we find the full ordered moment on the Co-site. The (002)  intensity provides us with a V moment M$_{\perp}$ = 0.419 ($\pm$ 0.067) $\mu_B$/V that represents the portion that is canted away from the collinear [$\bar{1}$$\bar{1}$0] direction, the ($\bar{2}$$\bar{2}$0) intensity confirms this value. If we assume that the (111) reflection is from a purely magnetic origin we obtain a collinear moment on the V-site of about 1.177 ($\pm$ 0.81) $\mu_B$/V.  Note that for V$^{3+}$ ordered moments between 0.6 - 1.3 $\mu_B$/V have been observed in various materials \cite{zhang} due to partial quenching of the orbital moment \cite{reig,zhang,LeeZnV,Reehuis2003,motome,garlea,sarkar,PLUMIER198753,ZhangCdV,mcqueen,mahajan,greg2012}. Utilizing the calculated perpendicular and parallel moments we find that the V-spins cant by $\sim$20$^{\circ}$ away from the [$\bar{1}$$\bar{1}$0] direction, but the spins remain in the $ab$-plane. Furthermore, we believe the small out of plane moment seen in magnetization (see Fig.\ref{MvsT}b) could be the result of applied fields overcoming the weak shape anisotropy for vanadium spins. %The in-plane magnetization values per unit cell (8 formula units) are smaller then expected if  3 $\mu_B$ on the Co-site and 0.7 $\mu_B$ on the V-site are used, this could be because the H = 0.1 T could be enough field to alter the in-plane canting angle.% For example, for Fe$_{1.18}$V$_{1.82}$O$_4$ values of $\sim$ 0.6 $\mu_B$ \cite{zhang} are reported, a similar value is found in ZnV$_2$O$_4$ \cite{LeeZnV,Reehuis2003,motome}, but larger values were found in MnV$_2$O$_4$ (1.3 $\sim$ $\mu_B$ \cite{PLUMIER198753,garlea,sarkar}), and CdV$_2$O$_4$ (1.2 $\sim$$\mu_B$ \cite{ZhangCdV}). The measurements in Fig. \ref{MvsT}b show low temperature magnetic moments per unit cell of about 5 $\mu_B$/u.c., when a $H$ = 0.1 T field is applied along the in-plane $a$ or $b$ axis, and about 1  $\mu_B$/u.c. when the same field is applied along the out-of-plane $c$-axis direction. Assuming for the moment that the reorientation indicates that the Co moment rotates fully into the $ab$-plane this residual out-of-plane moment should be solely from the V-site.

It is possible that we overestimate the parallel V moment by overlooking a possible intensity change of the (111) reflection due to structural distortions, i.e., movement of the oxygen atoms away from the bulk positions will increase the structural contribution to the (111) reflection. Note, if the parallel moment is indeed smaller the canting angle will increase since the intensity of the (002), and thus the size of the perpendicular moment, should not change due to a structural distortion involving the oxygen atom positions. The magnetic reorientation of the Co moments at 90 K, supported by both magnetization and neutron scattering measurements, indicates that a structural distortion likely accompanies this magnetic transition. This conclusion is indirectly supported by measured changes in scattering intensity at (115), (440) and (222) between high and low temperature by an amount which is too large to be explained by the spin ordering inferred from peaks at lower scattering angle. These changes are consistent with subtle displacements of oxygen anions, which would be difficult to observe with x-ray scattering but could lead to a modification of the single ion anisotropy of the Co atoms that would explain the strong drop in the ($\bar{2}$$\bar{2}$0) peak intensity and stabilize the observed low-temperature spin state. Similar shifts in oxygen anions have been observed in other orbitally ordered compounds \cite{Radaelli,Wheeler,Tokura462,ZhangCdV,PhysRevB.68.060405,motome,ishibashi,Mun,sarkar,Maitra,PhysRevLett.111.267201,PhysRevLett.94.137202}.  Note, there is a structural transition of the substrate at 105 K \cite{STO1,STO2} from cubic to tetragonal (associated with a change in the octahedral rotation); the temperature at which we observe magnetic phase transitions are far removed from this transition. 

We have grown epitaxial thin films of spinel vanadates. The strain-induced symmetry lowering and introduction of single ion anisotropy leads to dramatic differences in the low temperature non-collinear spin state compared to bulk. At the 90 K transition, a presumed structural distortion leads to a full reorientation of the Co moment from the [001] (c-axis, the easy axis in the collinear ferrimagnetic state) to the [110] direction in the film plane. Furthermore, the much larger perpendicular component of the V moment, compared to values for M$_{\perp}$ seen in powder samples \cite{reig}, indicates that the films display a canting angle of at least $\sim$20$^{\circ}$. This finding implies longer-ranged orbital order and an increased localized character of the V moments in strained CoV$_2$O$_4$ thin films relative to bulk samples. This work clearly shows that epitaxial strain is a viable knob to tune orbital order and itinerancy in spinel vanadates. More importantly, strain can be utilized in a wide variety of materials allowing more precise control of spin and orbital states, frustration, and itinerancy in thin films of frustrated antiferromagnets.   
%This implies more long-ranged orbital order and a push away from itinerancy is achieved in strained CoV$_2$O$_4$ thin films.
%\section{Acknowledgements}

A portion of this work was performed at the National High Magnetic Field Laboratory, which is supported by National Science Foundation Cooperative Agreement No. DMR-1157490, No. DMR-1644779, and the State of Florida. A portion of this research used resources at the High Flux Isotope Reactor, a DOE Office of Science User Facility operated by the Oak Ridge National Laboratory. Use of the Advanced Photon Source was supported by the U. S. Department of Energy, Office of Science, Office of Basic Energy Sciences, under Contract No. DE-AC02-06CH11357. GM and DR were supported by the National Science Foundation, under Grant No: DMR-1455264.

\bibliography{C:/Users/beekman/Documents/Proposals/DOE_career/career_Biblio}
% This file was created with JabRef 2.10.
% Encoding: Cp1252

\end{document}